# Assessment of bridge natural frequency as an indicator of scour using centrifuge modelling

KKGKD Kariyawasam(kdk26@cam.ac.uk), Campbell R Middleton(crm11@cam.ac.uk), Gopal Madabhushi(mspg1@cam.ac.uk), Stuart K Haigh(skh20@cam.ac.uk), James P Talbot(jpt1000@cam.ac.uk)
Department of Engineering, University of Cambridge, Cambridge, UK

ABSTRACT

One of the most prevalent causes of bridge failure around the world is "scour" – the gradual erosion of soil around a bridge foundation due to fast-flowing water. A reliable technique for monitoring scour would help bridge engineers take timely countermeasures to safeguard against failure. Although vibration-based techniques for monitoring structural damage have had limited success, primarily due to insufficient sensitivity, these have tended to focus on the detection of local damage. High natural frequency sensitivity has recently been reported for scour damage. Previous experiments to investigate this have been limited as a result of the cost of full-scale testing and the fact that scaled-down soil-structure models tested outside a centrifuge do not adequately simulate full-scale behaviour. This paper describes the development of what is believed to be the first-ever centrifuge-testing programme to establish the sensitivity of bridge natural frequency to scour. A 1/60 scale model of a two-span integral bridge with 15 m spans was tested at varying levels of scour. For the fundamental mode of vibration, these tests found up to a 40% variation in natural frequency for 30% loss of embedment. Models of three other types of foundation, which represent a shallow pad foundation, a deep pile bent and a deep monopile, were also tested in the centrifuge at different scour levels. The shallow foundation model showed lower frequency sensitivity to scour than the deep foundation models. Another important finding is that the frequency sensitivity to "global scour" is slightly higher than the sensitivity to "local scour", for all foundation types. The level of frequency sensitivity (3.1–44% per scour depth equivalent to 30% of embedment of scour) detected in this experiment demonstrates the potential for using natural frequency as an indicator of both local and global scour of bridges, particularly those with deep foundations.

# 1   Introduction

Bridge scour refers to the removal of soil from around structural foundations located in a river or coastal region as a result of the erosive action of water [1]. Scour around a bridge foundation leads to a reduction in bridge stiffness and stability because of the loss of embedment, which can even lead to bridge failure. According to historical bridge failure surveys, more than 50% of bridge failures have primarily been the result of scour-related causes [2, 3]. Monitoring bridge scour provides an opportunity to identify the potential risk of failure and to take pre-emptive countermeasures. However, detecting scour is more difficult than detecting superstructure damage since scour occurs underwater and is often not visible, and the harsh conditions during flooding can easily damage any underwater sensing equipment.

In contrast to other available techniques [4–8], vibration-based bridge scour monitoring is an indirect technique that does not require any underwater sensor installations. This method is based on the principle that scour causes a significant reduction in bridge stiffness, which leads to a measurable change in the natural frequencies of certain modes of vibration. Most of the previous research that focused on local crack detection rather than global damage such as scour has shown the natural frequencies of civil structures to be insufficiently sensitive to local structural damage. Even considerable damage, in the form of cracks as deep as half a bridge beam or pier, has, at best, indicated fundamental frequency sensitivity (0.4–7%) of the same order as environmental/operational sensitivity [9–14]. The damage considered in these previous studies has primarily been in the form of localised cracking, resulting in only local changes in stiffness and/or damping and therefore relatively small changes in the frequencies of global modes. Scour is a special damage case – effectively a change of boundary condition – that results in a global stiffness reduction and therefore significantly greater changes in natural frequency. A simple cantilever model illustrates this, in which the natural frequencies of a pier are inversely proportional to the square of the exposed length, although the soil fixity is a major simplification in this model and clearly needs to be considered carefully.

Vibration-based *scour* monitoring has shown considerable potential when analysed with computational models of bridge-soil systems [15–17]. The results of these numerical studies showed scour-induced changes in natural frequency as high as 30–40% for a 50% loss of pile embedment. A field study on a bridge, globally scoured by 3 m for repair purposes, indicated a 20% change in natural frequency [18]. The use of natural frequency to detect bridge pier integrity has been reported in Japanese railway bridges [19], although it has not been widely adopted in other parts of the world. A recent field deployment of this technique on a reinforced concrete road bridge indicated that the measurement error in natural frequency derived using ambient vibration data can be high, compared to the expected sensitivity of natural frequency changes due to scour [20–22]; therefore, it is important to experimentally establish the natural frequency sensitivity to scour of different bridges.

Establishing the sensitivity of natural frequency to scour in practice is difficult – a monitoring system would need to be installed on a candidate bridge, with no guarantee of measuring any scour within the timespan of the project. On the other hand, full-scale testing of controlled scouring of a bridge is not viable because of the costs involved, and tests on scaled-down soil-structure models at normal gravity [23, 24] do not simulate the real natural frequencies because of scaling issues, as elaborated in Section 2. Centrifuge modelling can be used to eliminate these scaling issues by allowing full-scale stress levels to exist within a small-scale model [25]. This technique has previously been used to examine the dynamic response of monopile models representing offshore wind turbines but not to examine the dynamic behaviour of bridges [26, 27].

Another important consideration is whether or not natural frequency is less sensitive to "local scour" at a bridge foundation than to "global scour". As shown in Figure 1, local scour refers to local lowering of the bed level relative to the general level of the channel, whereas global scour refers to a general lowering of the bed level over a wide area [28, 29]. The extended lowering of the soil level with global scour may result in a significant reduction in the stiffness of the underlying soil as the stiffness profile shifts down by the depth of scour. In contrast, local scour may not result in a significant reduction in the underlying soil stiffness due to the retention of overburden stress provided by the remaining soil surrounding the scour hole. As local scour would result in a smaller reduction in soil stiffness than global scour, bridge natural frequency may be less sensitive to local scour than to global scour.

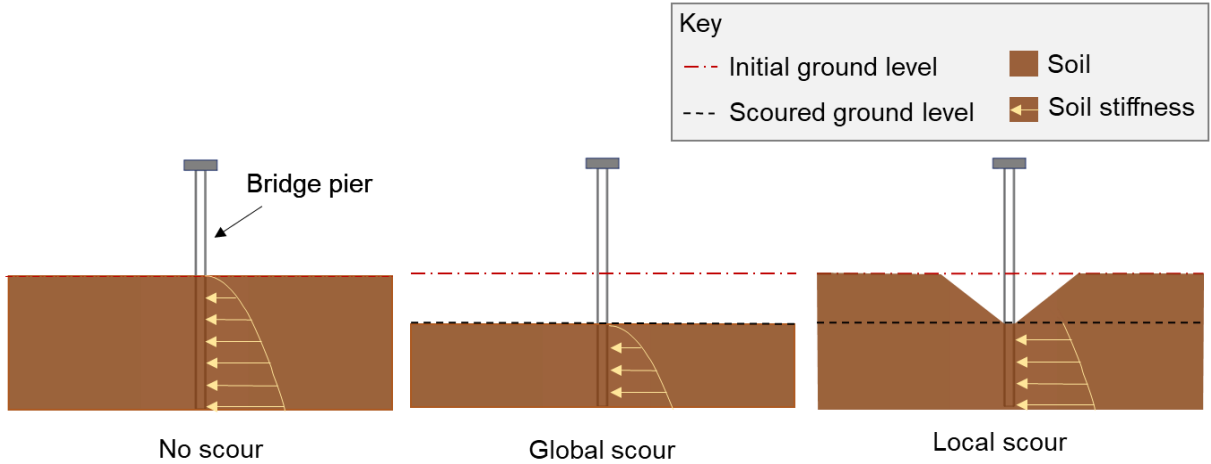

***Figure 1*** *Shape of a local and global scour*

This paper describes the development of a centrifuge experimental programme to capture natural frequency variation of bridges due to local and global scour. The objectives of the research were to:

(1) develop a scale-model testing methodology for assessing vibration-based scour monitoring techniques;

(2) identify the natural frequency sensitivity to scour of different bridge types;

(3) establish the natural frequency sensitivity of bridges due to different forms of scour, i.e. local and global scour.

## 2   Geotechnical centrifuge modelling

Centrifuge modelling is a technique used to correct inaccuracies in soil property scaling that is involved when attempting to model a full-scale (also called prototype-scale) soil mass with a small-scale (also called model-scale) soil model. To understand this scaling inaccuracy in the context of vibration behaviour applicable to this research, consider a 1/N scale model at normal gravity (1g) attempting to represent a full-scale soil mass, as shown in Figure 2.

Any full-scale depth $d_f$ is represented in the $1/N$ small-scale model by a corresponding depth of $d_f/N$. The full-scale self-weight stress ($\sigma_f$) at depth $d_f$ and the small-scale self-weight stress ($\sigma_s$) at the corresponding depth ($d_f/N$) are simply:

$$stress = \frac{force}{area} \qquad \sigma_f = \frac{d_f\, L^2 \rho g}{L^2} \;=\; \frac{d_f \rho g}{1} \qquad\qquad (1)$$

$$\sigma_s = \frac{\frac{d_f}{N}\left(\frac{L}{N}\right)^2 \rho g}{\left(\frac{L}{N}\right)^2} = \frac{d_f \rho g}{N} \qquad\qquad (2)$$

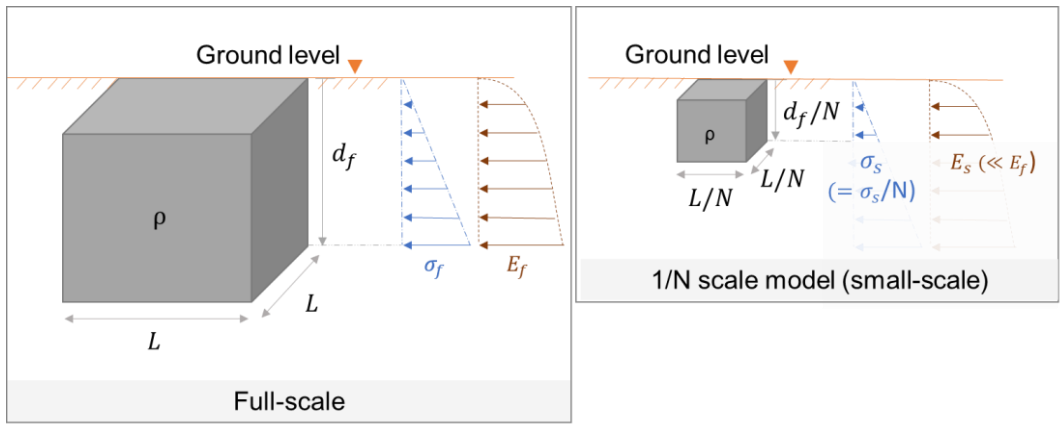

*Figure 2* Properties of full-scale and 1/N scale soil models (both at 1g)

Equations 1 and 2 give $\sigma_s = \sigma_f/N$, which suggests that the small-scale stress levels are 1/N times the full-scale stress levels at representative depths. The level of stress determines the stiffness properties of soil. For example, the small-strain elastic modulus of soil is directly related to the effective vertical stress, as given in Seed and Idriss (1970) [30] and illustrated in Figure 2. Therefore, in the same way as the small-scale stress level at $d_f/N$ is lower than the full-scale stress level at the corresponding location ($d_f$), the small-scale elastic modulus ($E_s$) at $d_f/N$ depth is smaller than the full-scale elastic modulus ($E_f$) at the corresponding full-scale depth ($d_f$). Hence, the elastic modulus profile of soil is different in the small scale and the full scale, and the dynamic behaviours in the two scales are not similar.

As soils exhibit highly non-linear stress-dependent behaviour, the correct behaviour will only be observed if the stress levels in the small and full scales match. Equal stress condition, that is, $\sigma_s = \sigma_f$, in equation 2 is attainable if the gravitational acceleration in the small scale is Ng instead of 1g. Centrifuge modelling does exactly this – it replicates full-scale stress fields within a small-scale model by increasing the effective gravitational field strength to compensate for the reduction in length, such that self-weight stresses are identical [25]. This necessitates testing a 1/N scale model at an effective gravity of Ng, which is generated by centrifuge rotation, such that the centripetal acceleration produces an increased vertical acceleration in the model.

Stress similitude between small and full scale at a length-scale factor of $N$ allows scaling laws to be derived for other parameters of interest such as flexural rigidity (scale factor $1/N^4$), mass ($1/N^3$) and frequency ($N$) [25]. To understand frequency scaling, consider, for example, a prismatic beam of length $L$, cross-sectional area of $A$, Young's modulus of $E$ and density of $D$. Based on the stiffness, mass and frequency relationship, the natural frequency ($f$) of the beam in the axial direction for undamped vibration is given by:

$$f = \frac{1}{2\pi}\sqrt{\frac{EA/L}{DAL}}$$

$$f = \frac{1}{2\pi L}\sqrt{\frac{E}{D}} \qquad\qquad (3)$$

According to equation (3), natural frequency is inversely proportional to the length, given that the elastic modulus and density of the material is the same. In centrifuge modelling, both the density and the elastic modulus of corresponding locations is maintained as constant despite the length scaling of $1/N$. Therefore, the natural frequency has a scale factor of $N$.

All tests for this research were conducted using the Turner Beam centrifuge at the Scofield Centre, University of Cambridge. This is a 10 m diameter centrifuge capable of subjecting models with a mass of up to 1000 kg to centripetal accelerations of 125g [31].

This centrifuge testing programme modelled the soil-structure interaction with a package of 434–485 kg in mass. It was tested under a 40g and a 60g effective gravitational field. Note that water was not explicitly modelled – see Discussion.

## 3  Full-scale and small-scale model element selection

These centrifuge tests were intended to model the change in dynamic response of various types of full-scale bridges using small-scale models. The constraints in the centrifuge model container and typical field conditions required iterative selection of the full-scale and small-scale properties. Only the finally selected properties and applicable constraints are given here.

### 3.1  Full-scale bridge selection

Three hypothetical full-scale bridges were selected to compare the viability of the vibration-based monitoring method in different types of bridges. The key differences and similarities among these bridges are listed in Table 1.

**Table 1** *The three types of full-scale bridges considered*

| Bridge | Deck type | Foundation type | Bridge deck arrangement | Spans | Material |
|--------|-----------|-----------------|-------------------------|-------|----------|
| 1 | Integral | Pile bent (deep) | Reinforced concrete composite deck with 0.2 m thick *in situ* slab on 8 Y1 precast beams | Two (15 m each) | Reinforced/prestressed concrete (C40) |
| 2 | Simply supported | | | | |
| 3 | | Pad (shallow) | | | |

The full-scale Bridge 1 was considered to be an "integral" bridge. Integral bridges have low maintenance costs because they have a monolithic connection at superstructure-substructure connections, without any bearings or expansion joints. Highways England recommends integral bridges as the first option to be considered for bridges with lengths not exceeding 60 m, skews not exceeding 30°, and settlements that are not excessive [32]. There are also a large number of integral bridges in the existing bridge stock (24% of EU bridges [33]). A flexible support abutment, which is a common abutment type used by bridge designers for integral bridges, was selected [34]. Flexible support abutments avoid any backfill interaction with the piles by adding piles through sleeves to create an annular void around the piles. This means that a full centrifuge model of this type of bridge does not require modelling of the abutment backfill.

The full-scale Bridges 2 and 3 considered were both "simply supported". A significant portion of the existing concrete bridge stock around the world also have simply supported bridge decks [33, 35]. Simply supported concrete bridges were constructed in the past because of their simplicity, ease of construction and ability to accommodate differential settlement of the supports. As a result of their many joints and bearings, this type of construction is no longer favoured in practice [36]. However, bridges with simply supported spans have been found to be the most susceptible to scour failure [37], perhaps because of the lack of redundancy compared to continuous and integral bridges.

The full-scale Bridges 1 and 2 were both selected to have pile bent foundations (a row of piles from deck level to the bottom of the foundation). This pile bent foundation arrangement included four piles of 18 m in length, 12 m of which was driven into the soil, with the remaining 6 m above ground level. The pier piles were assumed to be 760 mm in diameter. The abutment piles in Bridge 1, the integral bridge, were selected to be 540 mm in diameter. The abutments of Bridges 2 and 3, the simply supported bridges, were assumed to behave as fully rigid supports.

Bridge 3 was selected to have a shallow pad foundation. The shallow foundations are economically viable in bridges typically when the required soil properties are present within 3–4.5 m from ground level [38]. Hence, a shallow foundation depth of 3 m was assumed. The size of the shallow foundation was chosen as 4 m x 4 m in plan.

The maximum span of beam-slab-type bridges is between 10 and 30 m in the majority of bridges in Europe [33]. Therefore, a two-span bridge deck, with equal spans of 15 m, was chosen. The bridge deck was composed of 8 Y1 precast prestressed concrete beams, each with a second moment of area of $1.1 \times 10^{-2}$ $m^4$ and a cross-sectional area of 0.31 $m^2$. The spacing between Y1 beams was chosen as 1.5 m and the *in situ* slab depth was chosen as 0.2 m, based on the typical details given in a precast beam catalogue by Concast Precast Ltd (2009) [39]. A diaphragm beam of 1.85 m wide and 1.05 m deep was selected the locations where the pier and abutments meet the deck.

These full-scale hypothetical bridges were considered to be newly concreted bridges with a history of low stress levels, without significant cracking. Thus, all full-scale bridge elements constructed of C40 grade were assumed to have a modulus of elasticity of 35 GPa and a bulk density of 2550 $kgm^{-3}$ [40]. The effect of the higher elastic modulus of reinforcing and prestressing steel was ignored, since these contribute to only a small proportion (0.13–4%) of a typical cross-sectional area of a concrete element [41]. All foundations were assumed to be in a layer of uniform dense sand with 66% relative density.

## 3.2 Small-scale properties of bridge elements

Figure 3 (a) shows the scaling down of the full-scale Bridge 1. The full-scale bridge (with 15 m spans) was scaled down to a 1/60 scale (with 250 mm spans) to create Model 1 in the centrifuge container. Dynamic excitation for Centrifuge Model 1 was generated with an actuator. The mass of this actuator and other accelerometers (0.5 kg) at small scale was $1.1 \times 10^5$ kg in full scale, which was assumed to represent the extra mass of the three diaphragm beams in addition to the mass of the overlapping deck. The composite beam-slab deck in the full-scale Bridge 1 was simplified to a rectangular-slab bridge deck, having the representative flexural rigidity and mass in small-scale Model 1.

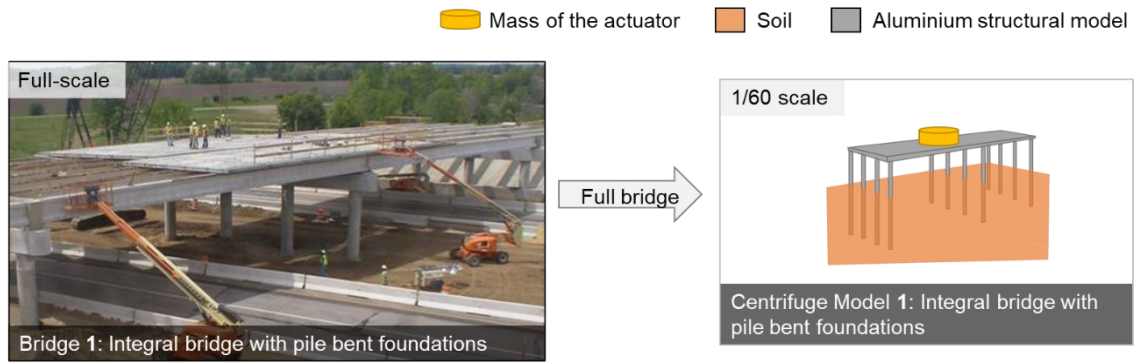

**(a)** *Full-scale Bridge 1 and small-scale Centrifuge Model 1*

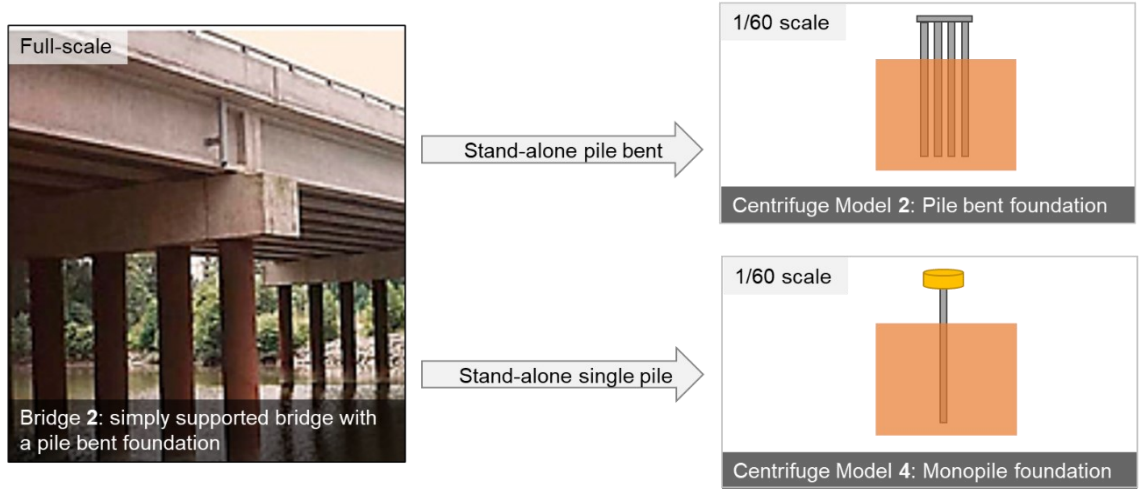

**(b**) *Full-scale Bridge 2 and small-scale Centrifuge Model 2 and 4*

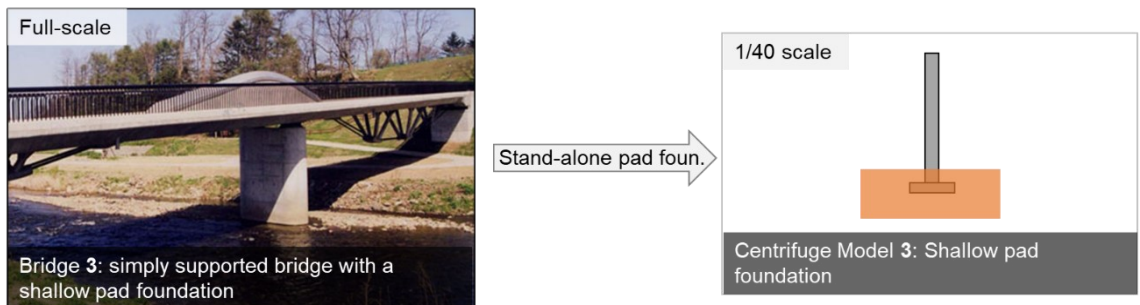

**(c)** *Full-scale Bridge 3 and small-scale Centrifuge Model 3*

Note: figures are not to scale, and the photographs are not the real full-scale bridges but only examples of similar types of bridges.

***Figure 3*** *Scaling down of three full-scale bridge types to four small-scale centrifuge models*

The presence of bearings in full-scale Bridges 2 and 3 complicates the testing of simply supported bridges in small-scale experiments. Therefore, only the "standalone foundations" (foundation-only models) of the bridge piers of Bridges 2 and 3 were scaled down as centrifuge models, with the aim of representing the local bending modes of the foundations that do not involve any deck vibration and thus can be represented by standalone foundations.

As shown in Figure 3 (b), Bridge 2, with a pile-bent-type pier foundation, was scaled down to a 1/60-scale standalone pile bent foundation model, Centrifuge Model 2. An individual pile foundation of the Bridge 2 pier foundation was scaled down to obtain an extra model, Centrifuge Model 4. Model 4 was excited with the same actuator that was used in Model 1. The small-scale Model 4, with its monopile and the combined mass of the actuator, top slab and piezoelectric accelerometer at the top (0.5 kg), was assumed to represent a full-scale monopile of 740 mm diameter with a mass of $1.1 \times 10^5$ kg at the top.

As shown in Figure 3 (c), Bridge 3, with a shallow pad foundation, was scaled down to a 1/40-scale standalone foundation model, Centrifuge Model 3. This foundation was scaled down by a factor of 40, instead of the factor of 60 used in the other models, to avoid the shallow depth of embedment in the small scale being too small in order to be able to accurately model for different scour depths.

All lengths of the elements were scaled down by $1/N$. The flexural rigidity ($EI$) and mass per unit length *($m_0$)* should be correctly scaled down by $1/N^4$ and $1/N^2$ respectively, to obtain the dynamic behaviour of the structural elements [25]. Furthermore, to obtain the correct soil-structure interaction, the foundation width should also be correctly scaled down by $1/N$.

All model structures were made from aluminium alloy elements. As a result of the change of material properties from full scale (concrete) to model (aluminium), not all of the centrifuge scaling laws could be met at the same time. Only the most critical parameters required were therefore selected to model the mass and stiffness properties needed to simulate the dynamic behaviour of the full-scale bridge. The correctly scaled critical properties are shaded in Table 2. For elements in soil (the piles and pad foundation), the most critical properties were selected as the flexural rigidity ($EI$) and the soil-structure dimensions. This selection required the piles to be thin hollow tubes, which leads to the mass per unit length of the model piles being lower than that required according to the scaling laws. Therefore, the hollow tubes were kept open-ended to enable the filling of soil mass into the piles as they were pushed into the soil model. For example, scaling laws require each 12.7 mm diameter pile to have a mass per unit length ($m_0$) of 0.31 kgm$^{-1}$, but the pile alone provides only 0.09 kgm$^{-1}$ (71% error), whereas the pile filled with soil provides 0.23 kgm$^{-1}$ (25% error). For the elements above ground level, such as the deck and columns, the flexural rigidity and mass per unit length were chosen as the main parameters for scaling, as there is no soil-structure interaction.

**Table 2** *Model element selection (shaded properties adhere to the scaling laws)*

| Bridge | Full-scale (reinforced/prestressed concrete) | | | | Model | Small-scale (aluminium) | | | |
|---|---|---|---|---|---|---|---|---|---|
| | Element | $EI$ ($Nm^2$) | $m_0$ ($kgm^{-1}$) | Soil-structure dimension (m) | | Element | $EI$ ($Nm^2$) | $m_0$ ($kgm^{-1}$) | Soil-structure dimension (m) |
| Bridge 1 | eight Y1 beams and a 0.20 m x 12.00 m top slab (composite deck) | $1.6 \times 10^{10}$ | 12444 | NA | Model 1 | one 12.7 mm x 100.0 mm rectangular solid section | 1195 | 3.43 | NA |
| | four 0.56 m diameter circular solid piles (abutment) | $1.7 \times 10^{8}$ | 628 | 0.56 | | four 9.0 mm outer diameter 0.9 mm thick wall circular hollow tube (abutment) | 13 | 0.06 (0.12 below ground level) | 0.009 |
| | four 0.74 m diameter circular solid piles (pier) | $5.2 \times 10^{8}$ | 1097 | 0.74 | | four 12.7 mm outer diameter 0.9 mm thick wall circular hollow tube (pier) | 40 | 0.09 (0.23 below ground level) | 0.0127 |
| Bridge 2 pier foundation | four 0.74 m diameter circular solid piles | $5.2 \times 10^{8}$ | 1097 | 0.74 | Model 2 | four 12.7 mm outer diameter 0.9 mm thick wall circular hollow tube piles | 40 | 0.09 (0.23 below ground level) | 0.0127 |
| | one 1.05 m deep 1.85 m wide rectangular solid capping beam | $6.2 \times 10^{9}$ | 4953 | NA | | one 12.7 mm x 40.0 mm beam rectangular solid beam | 478 | 1.37 | NA |
| Bridge 3 pier foundation | one 1.00 m x 0.70 m rectangular solid column | $1 \times 10^{9}$ | 1785 | 1 | Model 3 | one 12.7 mm x 32.5 mm rectangular solid column | 388 | 1.11 | 0.0325 |
| | one 1.27 m deep 4.00 m wide rectangular solid pad footing | $2.4 \times 10^{10}$ | 12950 | 4 | | one 25.4 mm deep x100 mm wide rectangular solid pad footing | 9560 | 6.86 | 0.1 |
| Bridge 2 monopile | one 0.74 m diameter circular solid pile | $5.2 \times 10^{8}$ | 1097 | 0.74 | Model 4 | four 12.7 mm outer diameter 0.9 mm thick wall circular hollow tube | 40 | 0.09 (0.23 below ground level) | 0.0127 |
| | one 1.05 m deep 1.85 m wide rectangular solid slab section | $6.2 \times 10^{9}$ | 4953 | NA | | 12.7 mmx 40.0 mm rectangular solid slab section | 478 | 1.37 | NA |

## 4    Centrifuge experimental programme

Based on the small-scale properties derived in the previous section, Models 1–4 were constructed in a soil medium in a cylindrical centrifuge container. Figure 4 and Figure 5 show a diagram and photograph of the experimental setup. All model foundations were kept at least 100 mm away from the centrifuge container wall and 50 mm away from the bottom of the container to limit boundary effects on the soil-structure interaction [42]. All four models were placed adjacent to one another in the cylindrical centrifuge container, and only two models could be tested simultaneously during each centrifuge test flight because of the number of excitation sources available. The preparation of the structural models, the soil and the full experimental set-up is discussed in detail in the following sections.

### 4.1    Model structure preparation

The model structures were made from aluminium sections and had the small-scale properties listed in Table 2. Aluminium sections were used instead of the full-scale material, reinforced/prestressed concrete, as aluminium sections can be machined and fabricated to a higher degree of accuracy at small scale than it is possible to do with concrete. Paper rulers were glued on to all model foundations with the aim of using them as a reference level when creating scour holes (Figure 8).

### 4.1.1    Material properties

The circular hollow piles in the model structures were made of aluminium alloy 6061-T6, and the rest of the solid sections were made of aluminium alloy 6082-T6. These aluminium alloys have Young's modulus of 70 GPa and density of 2700 kgm$^{-3}$ [43, 44].

### 4.1.2    Connections

All of the model structures had at least one integral connection, either at the foundation to deck connection or at the stub column to pad foundation connection. To obtain these integral connections, sockets of the same cross-sectional area as the column/pile elements were first machined up to half the depth of the slab/pad section. The columns/piles were then inserted into these sockets and glued in place with a high-strength retainer adhesive. The connection rigidity of each pile/column was tested by finding the fundamental sway natural frequency of the piles when the slab/base was fixed. The piles without adequate integral connection showed a varying frequency for repeated tests or a lower frequency than other similar piles. These piles were removed and reconnected to obtain the desired integral connection.

### 4.1.3    Instrumentation

The accelerometer locations on Model 1 are shown in Figure 4 (a) and on Models 2, 3 and 4 are shown in Figure 4 (b). The accelerometer arrangement was chosen to capture the fundamental sway mode and local pier modes of vibration. Piezoelectric accelerometers and smaller size MEMS accelerometers were mounted on the decks and piles. The piezoelectric ("P") accelerometers were DJB A/23/TS accelerometers, with a working frequency range of 10–10,000 Hz [45]. The ADXL78 ("m") accelerometers claim 1100 µg/√Hz noise density for 10–400 Hz frequency range and a range of 35g [46]. The ADXL1002 ("M") accelerometers claim far less noise density of 40 µg/√Hz for 1–10,000 Hz frequency range and a range of 50g [47]. A high acceleration range was chosen since horizontal accelerometers, even with a slight tilt, may measure part of the effective gravity field (40–60g) in the centrifuge.

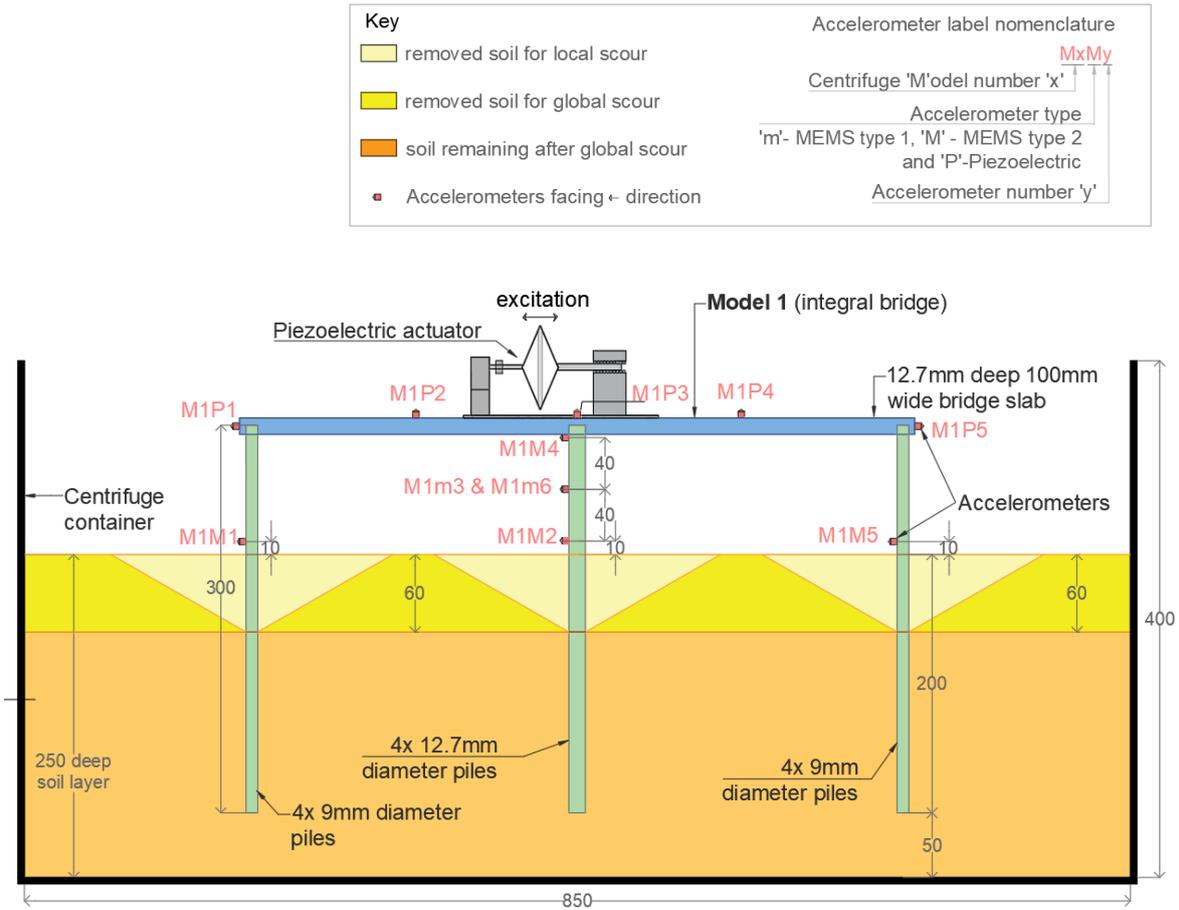

**(a)** *Centrifuge Model 1*

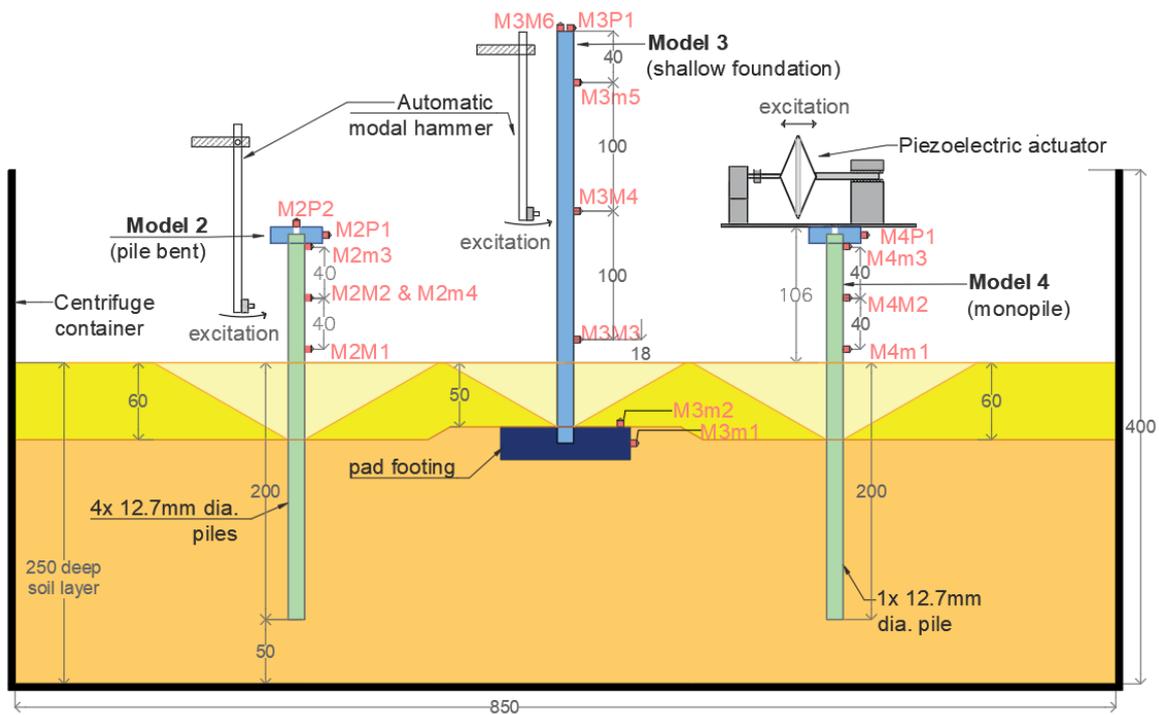

**(b)** *Centrifuge Models 2, 3 and 4*

**Figure 4** *Experimental set-ups, sensor arrangement and scour levels corresponding to all the models in the centrifuge container – all dimensions are in mm*

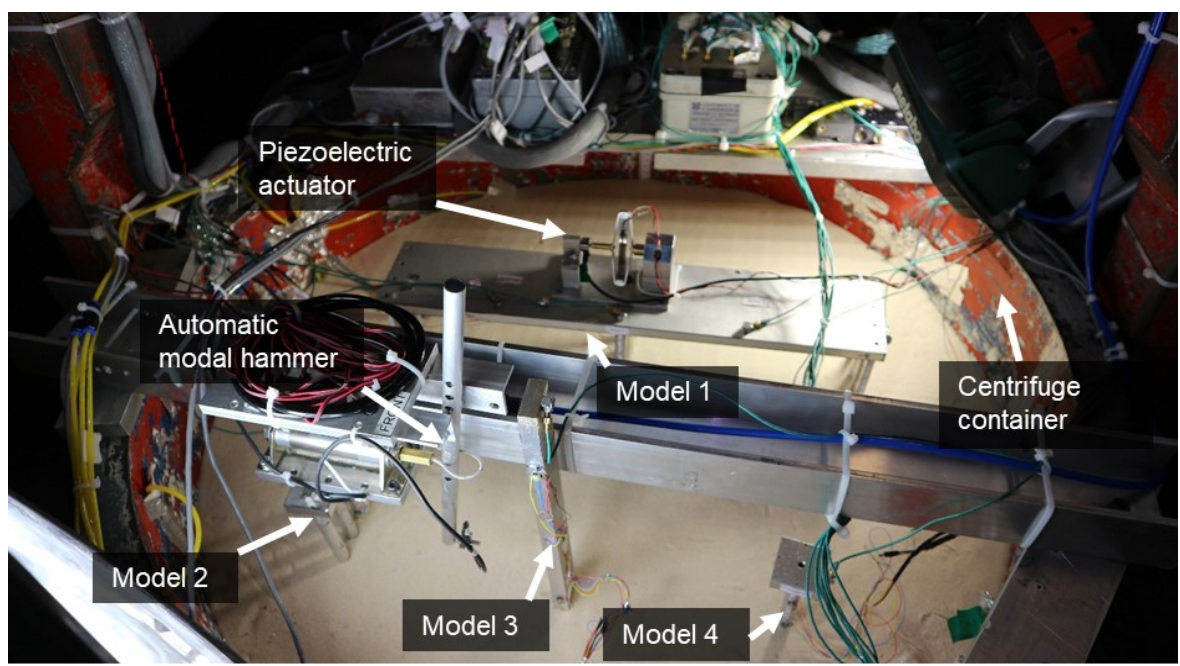

***Figure 5*** *The cylindrical centrifuge container with all four structural models and the excitation set-up*

## 4.2   Soil model preparation

The soil model was prepared in a cylindrical steel container, as shown in Figure 5. The cylindrical centrifuge container was 850 mm in diameter and 400 mm in depth and filled with a 250 mm deep layer of sand.

### 4.2.1   Material properties

The soil used for the centrifuge test was Hostun sand acquired from Drôme in the south-east of France. This sand type is widely used in centrifuge model tests. It is a high silica ($SiO_2 >$ 98%) sand of grain shape varying from angular to sub-angular [27]. The properties of Hostun sand are given in Table 3.

***Table 3*** *Geotechnical properties of Hostun sand* [27]

| Property | Value |
|---|---|
| $D_{10}$ | 0.286 mm |
| $D_{50}$ | 0.424 mm |
| $e_{min}$ | 0.555 |
| $e_{max}$ | 1.010 |
| $G_s$ | 2.65 |
| $\Phi_{crit}$ | 33° |

### 4.2.2   Sand pouring

The sand was poured with an automatic sand pourer developed by Madabhushi et al. at the Scofield Centre centrifuge testing facility [48]. This pourer allows the preparation of a sand sample of uniform density in the centrifuge container by maintaining the same drop height.

After a calibration sand-pouring test, a drop height of 690 mm through a 5 mm diameter nozzle (no sieve), with a 20 mm spacing between consecutive motions, was selected to obtain a relative density of 66%. The relative density of the final sand model in the centrifuge container was found using the pre-post weight difference and the properties given in Table 3. The bulk density of the soil was found to be 1550 kgm$^{-3}$.

The sand pouring was paused when the sand fill was at 175 mm to place Model 3 with the shallow foundation, and then the pouring continued to embed the foundation.

The structural models with pile foundations (Models 1, 2 and 4) were inserted into the soil model at the end of the sand pouring. The piles were pushed down until the desired embedment depth of 200 mm had been reached. It was assumed that the piles with open ends would not cause significant disturbance to the soil when inserted at normal gravity.

## 4.3  Excitation sources

Two excitation sources were used to test two models during each centrifuge flight. A piezoelectric actuator, which had previously been used to test monopile foundations [27], excited Models 1 and 4, and a newly developed automatic modal hammer excited Models 2 and 3.

### 4.3.1  Amplified piezoelectric actuator (APA)

The amplified piezoelectric actuator used was an APA400MML [49]. As shown in Figure 6, this actuator has one end fixed to its base through a load cell and the other end connected to a brass mass on roller bearings. Excitation of the piezoelectric material transfers a horizontal inertial force to the APA base, which was fixed on the top of Models 1 or 4.

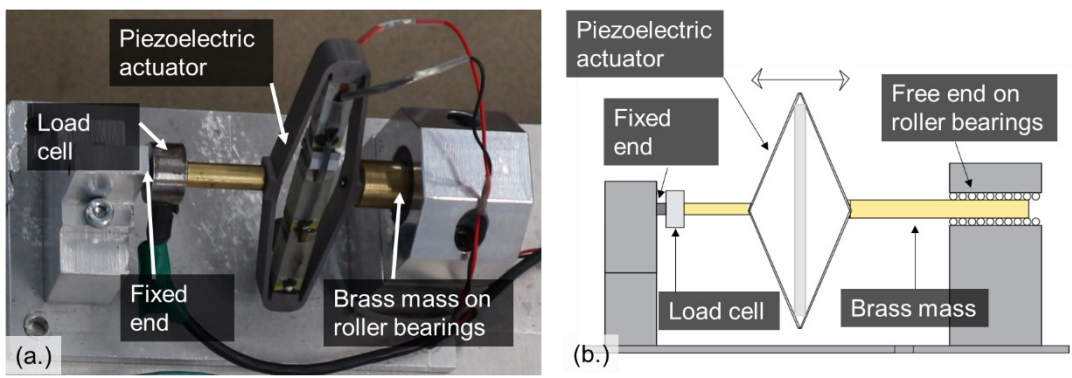

***Figure 6*** *Amplified piezoelectric actuator set-up (a) in the experiment (b) diagram*

### 4.3.2  Automatic modal hammer

An automatic modal hammer was developed to excite Models 2 and 3, as it was thought to be able to provide a better external excitation than the piezoelectric actuator, which rests on the model. As shown in Figure 7, impulsive excitation was provided by a free-hanging kicker made from an aluminium tube with a load cell at the bottom end. An air hammer, with two high-pressure airlines controlled through a relay switch, was used to provide an impulse to the kicker. The impulse was transmitted to the model through the load cell, which measured the force input to the model. A voltage input to the relay automatically excited the modal hammer every 2 s.

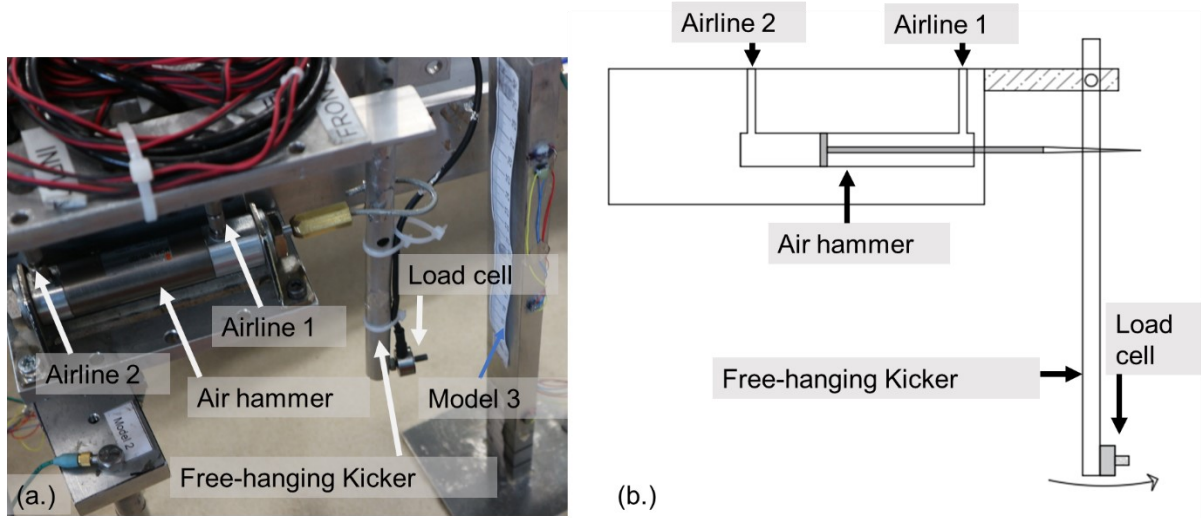

**Figure 7** *Set-up of the automatic modal hammer controlled by air-pressure (a) in the experiment (b) diagram*

## 4.4 Data acquisition system

All data acquisition was carried out with the centrifuge data acquisition system, which uses slip rings to transfer data to the computers in the centrifuge control room. All signals were transmitted through a set of amplifiers and then through an analogue to digital converter [25]. The digitalised signals were logged by DasyLab software at a sampling rate of 10 kHz.

## 4.5 Centrifuge flight programme

Before each centrifuge flight started, the desired scour hole was created using vacuum suction in order to limit disturbance to the underlying soil. The depths of the scour hole were measured with reference to the paper rulers that were on all of the structural model foundations. Figure 8 (a) and (b) show the maximum local and global scour holes. With global scour, the soil surface level was maintained horizontal. With all local scour holes, an approximate inclined angle of $30^0$ was maintained around all sides (Figure 8), which was slightly below the angle of the response of Hostun sand ($33^0$). The typical slope of a local scour hole is equal to the angle of response of the bed material for the upstream side but could be lower than that for the downstream side [50].

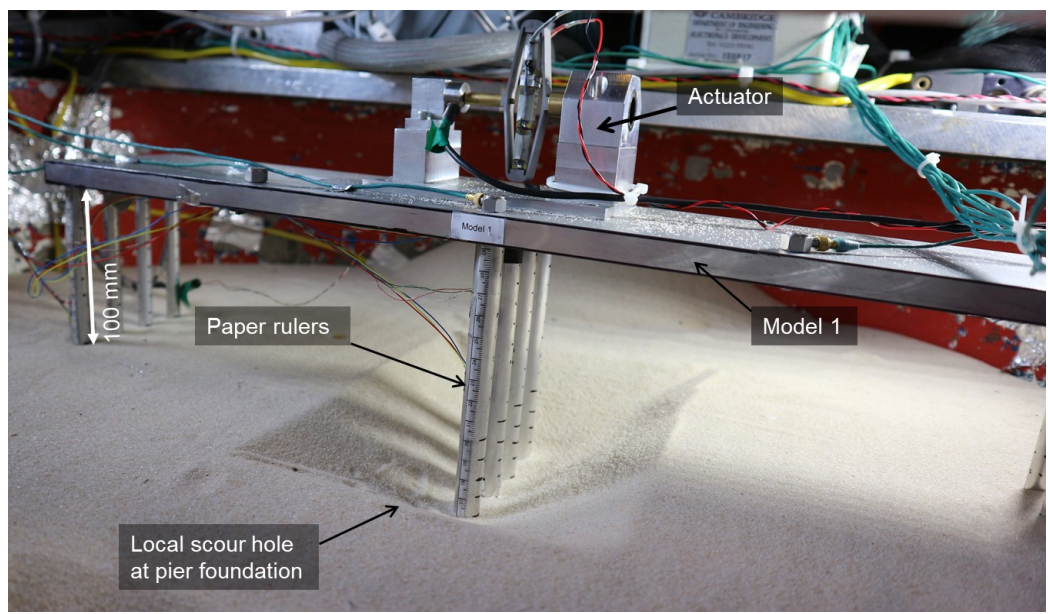

*Figure 8 A local scour hole created at the pier foundation of Model 1*

Following the creation of the scour holes, the natural frequencies of the models were measured at normal gravity, 1g, to confirm the operation of the excitation sources and data logging before the centrifuge flights were started. Then, the centrifuge beam rotation was initiated. When the rate of rotation of the beam provided the desired centripetal acceleration (40g or 60g) to the models, the natural frequency tests were conducted again. Finally, the centrifuge flight was stopped, and the next scour step started.

The different steps of scour tested around each centrifuge model are summarised in Table 4. Model 1 excitation was provided by the piezoelectric actuator for the cases simulating local scour around the middle pier, and also for the final global scour case. Some additional tests were conducted with ambient vibrations alone in Model 1, while Model 4 was being tested with the piezoelectric actuator. To maintain the same mass as Model 1 in all tests, the mass of the piezoelectric actuator removed from Model 1 during these additional tests was replaced by a standard metal mass plate of 0.5 kg. There was an unintended error here since the mass of the actuator was 0.44 kg, not 0.5 kg. This error in the additional tests, however, was found not to have affected the natural frequency observations, when M1S7a and M1S7b with same scour depths were compared (see Section 6).

With Model 3, local scouring was introduced in 17 mm steps (0.68 m in the full scale), with a final 50 mm (2 m) global scour case, requiring a total of five centrifuge flights. With Models 2 and 4, local scouring was introduced in two 30 mm steps (1.8 m) up to 60 mm (3.6 m), and a final global scour of the same depth was tested, making a total of four centrifuge flights.

The air temperature during the one-week centrifuge flight programme was measured at the end of every centrifuge flight. The temperature was stable (12.5–16 $^0$C) throughout the experimental programme, as the centrifuge is located underground. Therefore, it was assumed that there were negligible temperature effects on the natural frequencies of the models, and they varied entirely as a result of scour.

*Table 4* The scour steps around the bridge models

| Model | Step name | Centrifuge flight number | Equivalent full-scale scale scour case | Location of scour for each step | Excitation source |
|-------|-----------|--------------------------|----------------------------------------|----------------------------------|-------------------|
| 1 | M1S1 | 1 | no scour | Middle pier only | Piezoelectric actuator |
| | M1S2 | 2 | 1.2 m local scour | | |
| | M1S3 | 3 | 2.4 m local scour | | |
| | M1S4 | 4 | 3.6 m local scour | | |
| | M1S5 | 6 | 3.6 m local scour | + Left abutment | Ambient vibration |
| | M1S6 | 7 | 3.6 m local scour | + Right abutment | |
| | M1S7a | 8 | 3.6 m global scour | Everywhere | Ambient vibration |
| | M1S7b | 9 | | | Piezoelectric actuator |
| 2 | M2S1 | 5 | no scour | Pier | Automatic modal hammer |
| | M2S2 | 6 | 1.8 m local scour | | |
| | M2S3 | 7 | 3.6 m local scour | | |
| | M2S4 | 8 | 3.6 m global scour | | |
| 3 | M3S1 | 1 | no scour | Column | Automatic modal hammer |
| | M3S2 | 2 | 0.68 m local scour | | |
| | M3S3 | 3 | 1.36 m local scour | | |
| | M3S4 | 4 | 2.00 m local scour | | |
| | M3S5 | 9 | 2.00 m global scour | | |
| 4 | M4S1 | 5 | no scour | Pile | Piezoelectric actuator |
| | M4S2 | 6 | 1.8 m local scour | | |
| | M4S3 | 7 | 3.6 m local scour | | |
| | M4S4 | 8 | 3.6 m global scour | | |

## 5   Fixed-base tests

For each scour level, the soil around the piles provides some stiffness to the models. The maximum soil stiffness, and thus the maximum natural frequency, is reached when the embedded layer of soil provides full fixity. Therefore, as shown in Figure 9, the models were tested with fixity in order to obtain an upper-bound natural frequency to help confirm the centrifuge test results. The exposed height of the models was increased to simulate the scour depths. It is noteworthy that the fixity was provided using clamps and therefore the ideally fixed condition was not attainable.

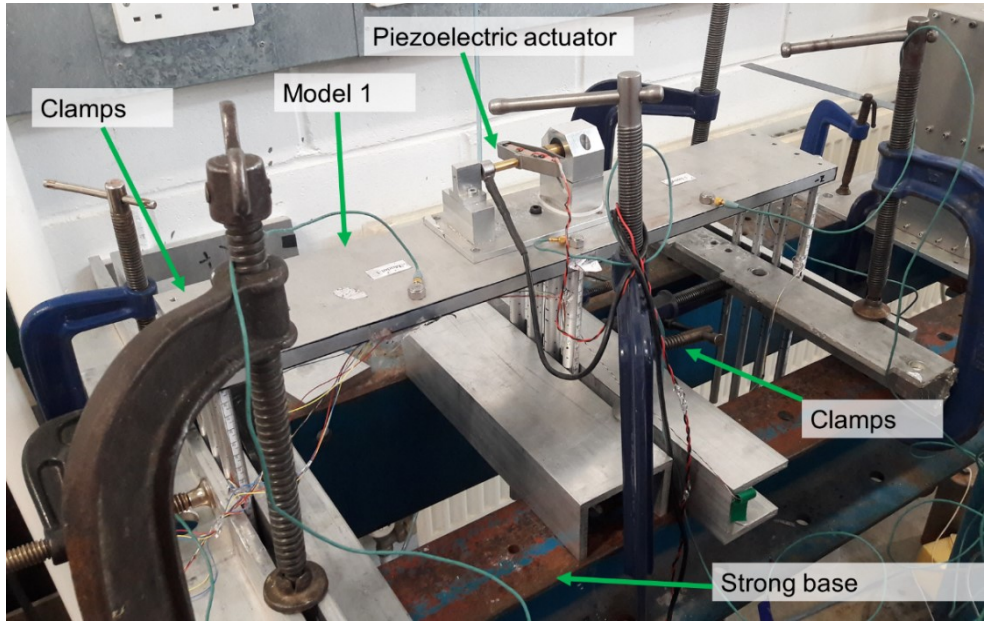

**Figure 9** *Fixed-base test being carried out for Model 1*

## 6 Modal analysis

The logged input and output data were analysed in MATLAB. An example input excitation, and the corresponding response measured in Model 1, are shown in Figure 10. The transient response to the APA was of insufficient duration to determine the frequency content with acceptable resolution. However, there was sufficient ambient excitation between the APA impulses for the required determination of natural frequency.

Modal analysis for Models 1 and 4 was therefore carried out using the output-only method, frequency domain decomposition (FDD). The FDD method is based on the singular value decomposition of the measured PSD matrix $G_{yy}(f)$ at discrete frequencies $f$, as shown in Equation (4), where $U(f)$ is the matrix of singular vectors and $S(f)$ is the diagonal matrix of singular values. Over the frequency range associated with a peak in the first singular values, the structural response is dominated by a single vibration mode, with the first singular vector being an estimate of the mode shape, and the corresponding first singular value being the auto-PSD of the modal contribution. The FDD method assumes that the input excitation is wideband (white noise) and that the structure is lightly damped [51, 52].

$$G_{yy}(f) = U(f)[S(f)]U(f)^T \qquad (4)$$

Figure 11 shows the first singular value spectra of the Model 1 outputs measured in two centrifuge flights, both having a global scour depth of 3.6 m at full scale. One centrifuge flight had Model 1 excited purely by ambient vibration, while the other had the piezoelectric actuator as well. Both flights indicated a sway mode shape of the bridge (Figure 12) at the same frequency (1.1 Hz in full scale), which indicates that the ambient vibration alone enables repeatable detection of natural frequency.

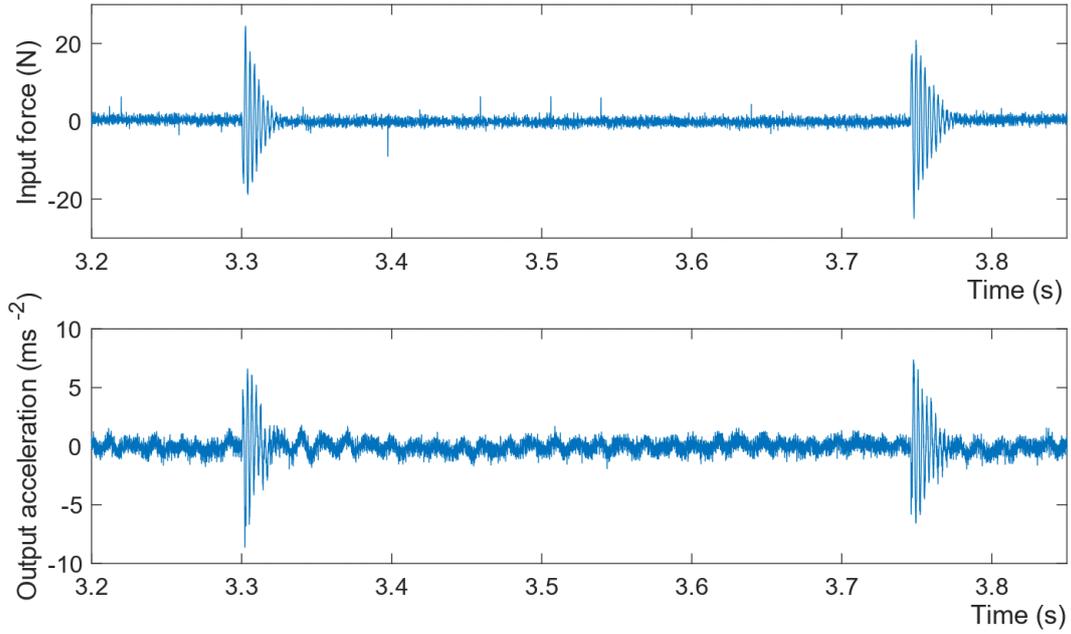

**Figure 10** *Measured input excitation and the response acceleration of one sensor in Model 1 (scour step M1S7b in Table 4)*

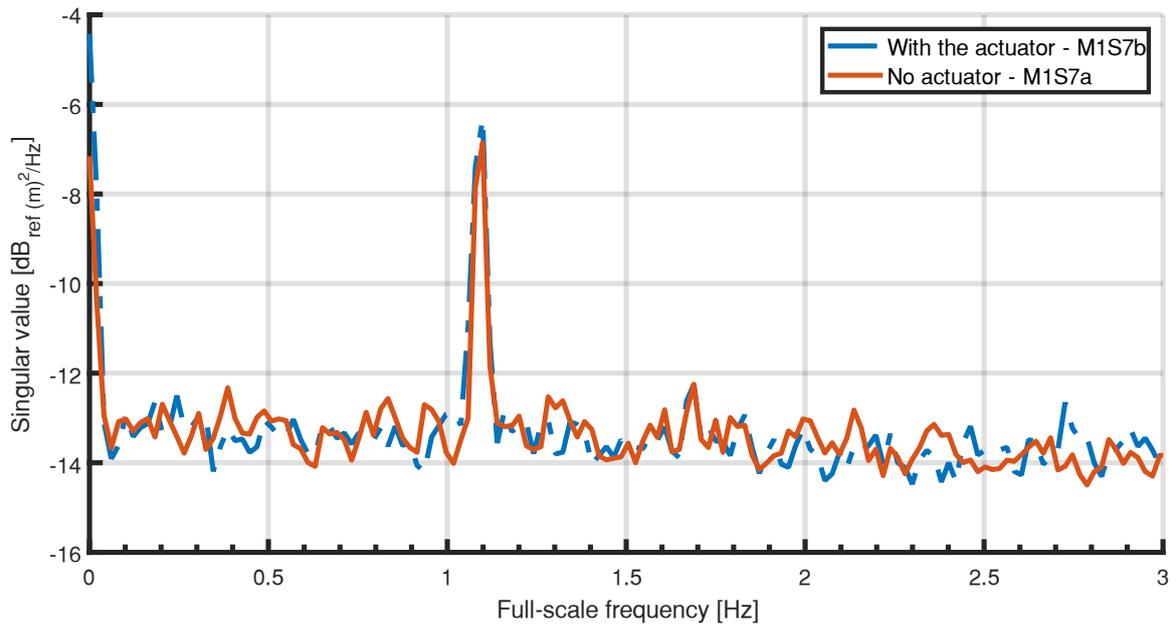

**Figure 11** *First singular value of Model 1 response for the same scour case but with two excitation sources: no actuator (step M1S7a), and with actuator (step M1S7b)*

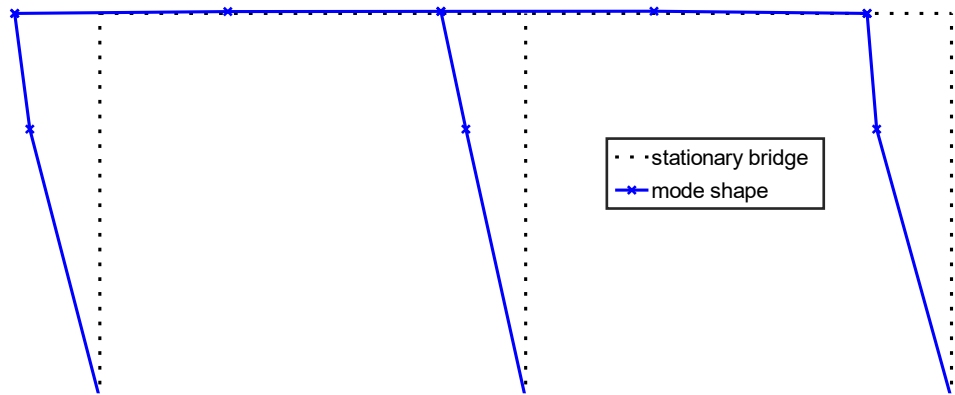

**Figure 12** *A typical sway mode shape of Model 1 – side view ("X" – measurement locations)*

Models 2 and 3 were excited by impacts from the automatic modal hammer, and the natural frequencies found from a simple frequency-response function (FRF). The FRF quality was improved by segmenting the signals by individual impacts and applying a force window to each segment of input and an exponential window to each segment of output [53], as shown in Figure 14. An example acceleration FRF magnitude and coherence are shown in Figure 13. Peaks in the FRF correspond to vibration modes, the first of which in this case is 2.22 Hz.

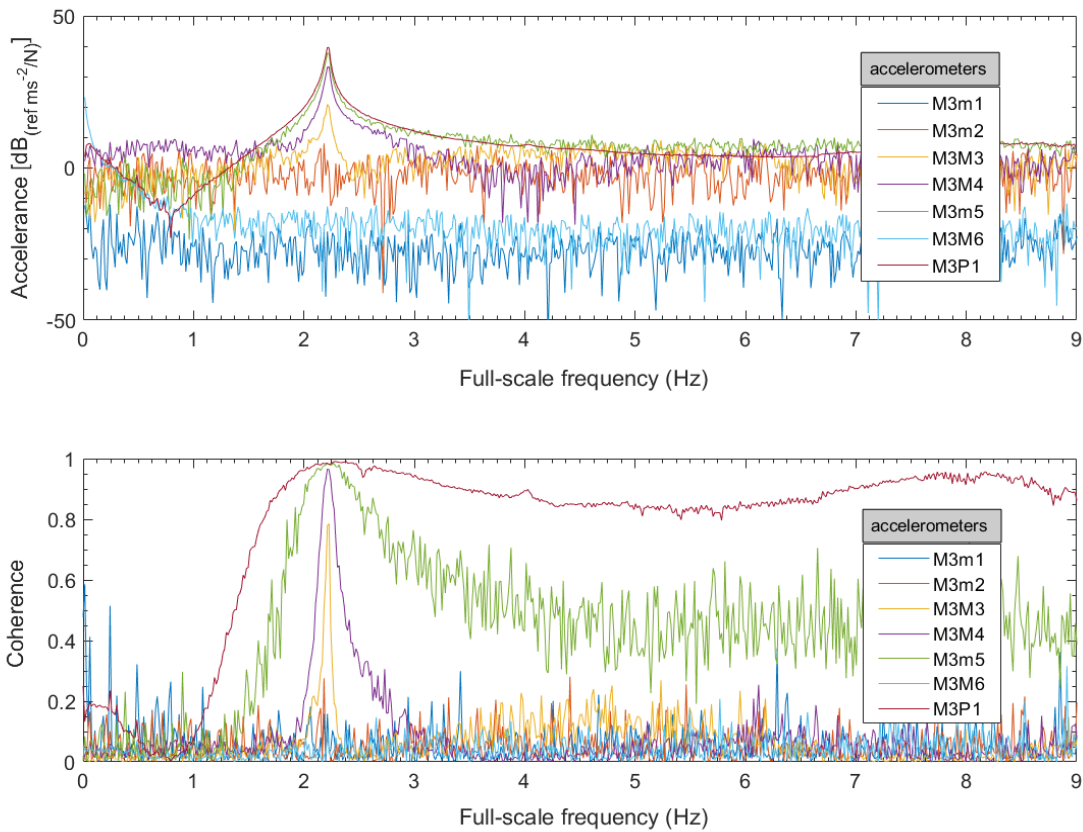

**Figure 13** *Acceleration FRF and coherence for Model 3 (scour step M3S5 in Table 4)*

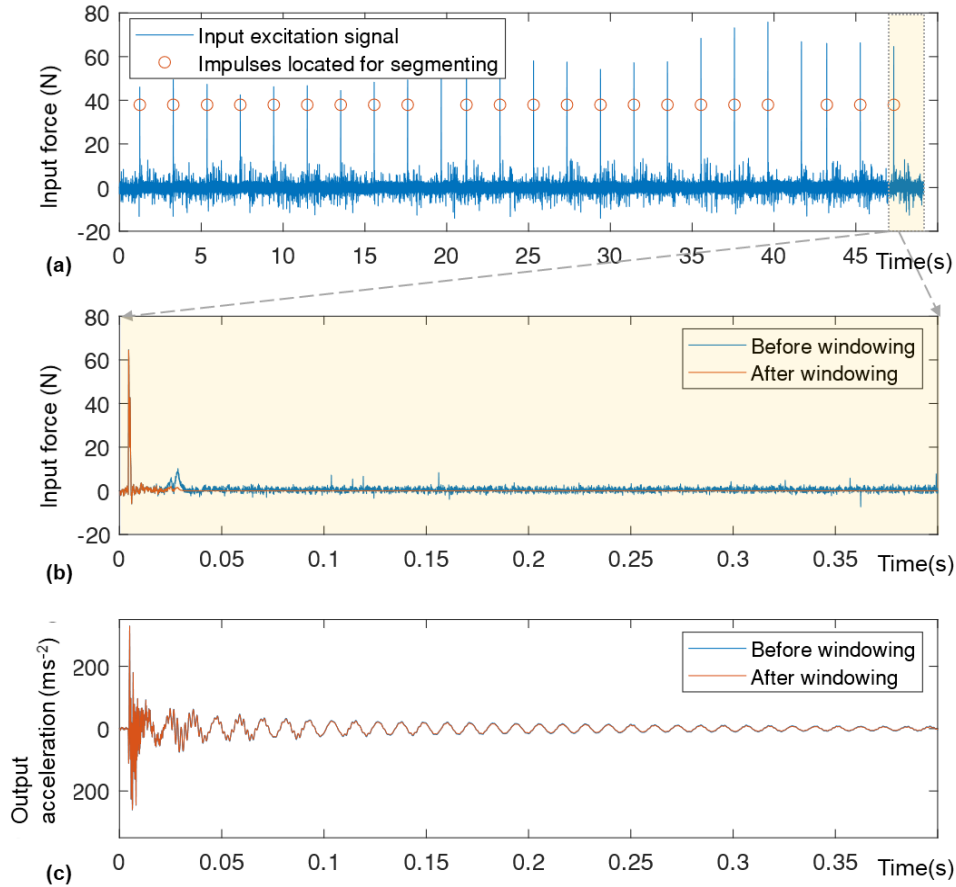

***Figure 14*** *Time histories of Model 3 – scour step M3S5 (a) input force, (b) part of a segmented input force, and (c) corresponding output acceleration of M3P1*

# 7   Natural frequency variation

As the environmental parameters were controlled, any change in natural frequency can be attributed to the scour itself. Therefore, high sensitivity of natural frequency indicates a high potential for it to be used as an indicator of scour, while low sensitivity indicates otherwise.

The spectra obtained from modal analysis showed reliable modal peaks only for the fundamental modes, and thus all of the natural frequencies discussed here correspond to the fundamental sway mode (Figure 12). All frequency measurements have been scaled to full scale using the corresponding scale factors (40 for Model 3 and 60 for all other models).

For comparison, the natural frequencies measured at "1g" in the centrifuge container with soil, and in the "fixed-base" condition, are also plotted. At "1g", the soil was less stiff than the soil during the centrifuge test, and therefore the "1g" frequency measurements give a lower bound for the frequencies measured during the centrifuge test. The fixed-base condition represented the maximum stiffness that the soil can provide, and it therefore provides an upper bound.

## 7.1   Model 1: Bridge 1 (integral bridge)

The full-scale natural frequency variation found using the integral bridge model is shown in Figure 15. As expected, the natural frequency measurements at 60g remained between the fixed-base upper bound and 1g lower bound at all scour levels. The spectral plots of all scour cases showed a noise peak at 1.67 Hz (small-scale 100.1 Hz), which is likely to be the result of either electrical or mechanical noise in the centrifuge. This overshadowed the natural frequency measurement of step M1S2 in Table 4 (1.2 m of local scour at the middle pier tested at 60g).

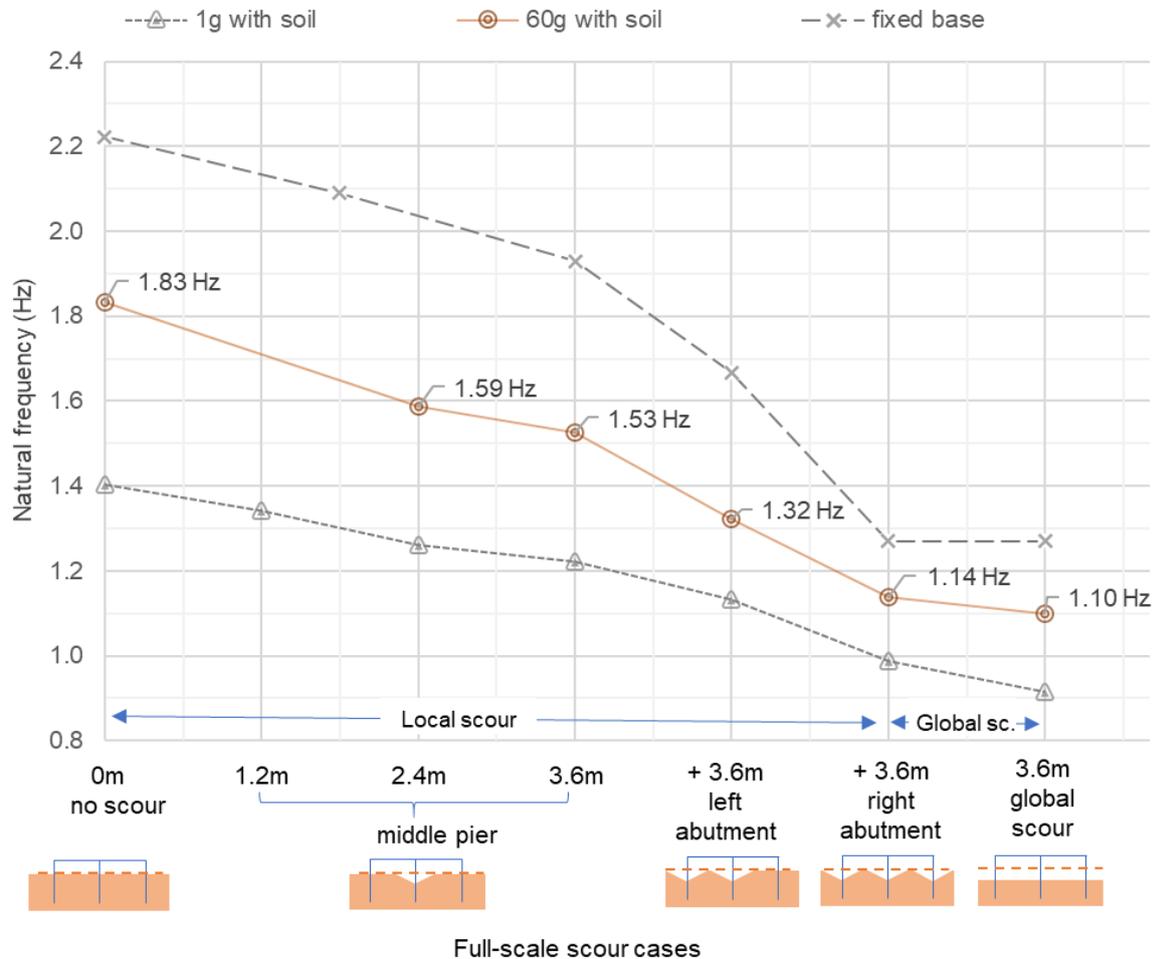

***Figure 15*** *Representative full-scale natural frequency variation of integral bridge*

### 7.1.1 Local scour

As shown in Figure 15, 3.6 m of local scouring at the middle bridge pier foundation resulted in a 16.4% reduction in the natural frequency. This corresponds to a frequency sensitivity of approximately 4.6% per metre of full-scale scour.

An additional 3.6 m of local scouring at the left abutment caused a frequency reduction of 0.21 Hz (-14%) from 1.53 Hz. Furthermore, 3.6 m of local scouring at the right abutment resulted in a 0.18 Hz (-14%) reduction in natural frequency from 1.32 Hz. Both of these local scour cases at the abutments represent approximately 4% of natural frequency shift for every metre of full-scale local scour at one of the abutments. The cumulative effect of 3.6 m deep local scour at each of the three supports resulted in an overall 38% reduction in natural frequency.

### 7.1.2   Global scour

As shown in Figure 15, 3.6 m of full-scale global scour represents a 40% change in natural frequency, which corresponds to a frequency sensitivity of approximately 11% per metre of global scour. There was only a small difference between the frequency sensitivities to global and local scour at all foundations. The overall change in natural frequency due to local scour at all foundations was 38%, whereas global scour of the same depth represents a 40% change. Therefore, some effect due to the additional confining pressure remained during local scour, but this only caused a 0.5% difference in frequency sensitivity per metre of full-scale scour depth.

## 7.2   Model 2: pile bent foundation of Bridge 2

The full-scale natural frequency variation of the pile bent foundation is shown in Figure 16 – as expected, all of the natural frequency measurements at 60g fall between the fixed-base upper bound and 1g lower bound for all scour levels

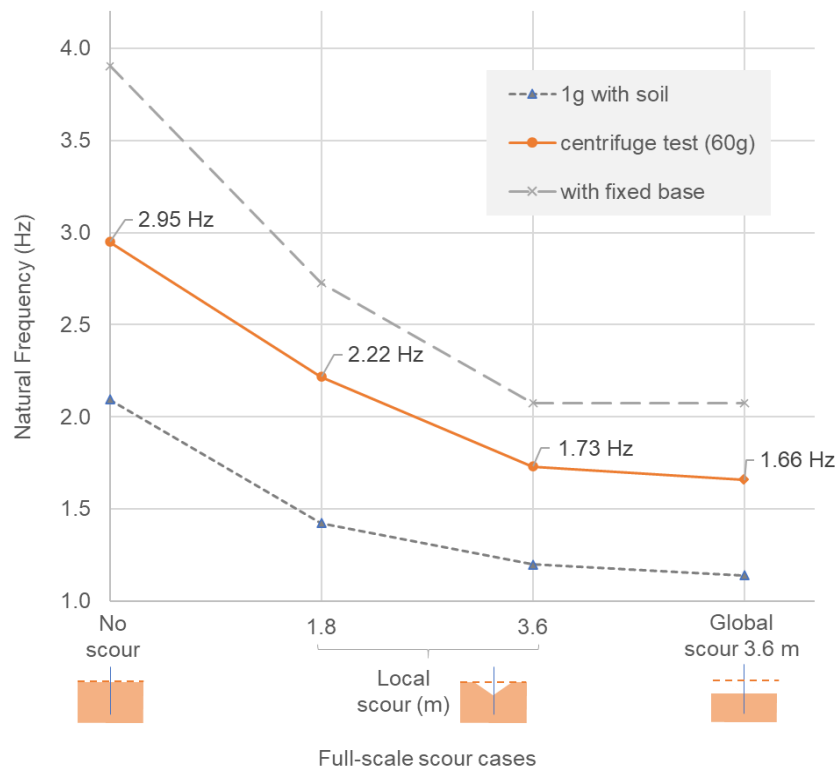

***Figure 16** Full-scale natural frequency variation of pile bent*

### 7.2.1   Local scour

The initial 1.8 m of local scour caused the natural frequency at 60g to fall from 2.95 Hz to 2.22 Hz, which represents a 25% reduction in natural frequency. A further 1.8 m of scour resulted in a 22% change, which indicates that the natural frequency sensitivity to local scour reduced slightly as the scour deepened.

The total 3.6 m of local scour gave a 41.4% frequency reduction. This reduction corresponds to an average natural frequency sensitivity of 11.5% per metre of full-scale local scour.

### 7.2.2  Global scour

The global scour of 3.6 m caused the natural frequency of the model to fall to 1.66 Hz, lower than the representative local scour value of 1.73 Hz. However, the overall natural frequency sensitivities due to the local and global scour were not significantly different. The overall natural frequency sensitivity to global scour of 3.6 m was 43.7%, and due to local scour of 3.6 m was 41.4% – only a 0.6% sensitivity difference due to 1 m of scour.

## 7.3  Model 3: shallow foundation of Bridge 3

The variation of the natural frequency with scour depth of the shallow foundation is shown in Figure 17. All of the natural frequency measurements at 40g lie below the fixed-base upper bound line. The modal analysis conducted for the 1g vibration measurements did not show modal peaks in the spectra, and thus natural frequency could not be retrieved, possibly as a result of insufficient fixity being provided by the shallow depth of soil at 1g.

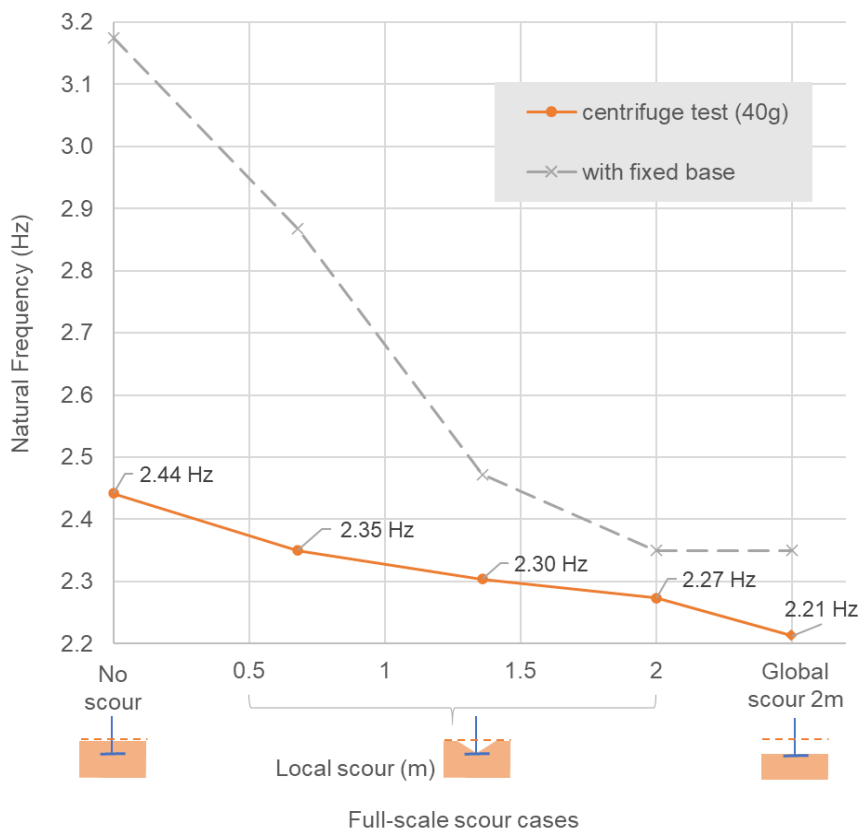

***Figure 17*** *Full-scale natural frequency variation of shallow foundation*

### 7.3.1  Local scour

A local scour hole of 2 m at full scale was created in three equal steps around the column of the shallow foundation. The full 2 m of local scour caused the original frequency of the model (2.44 Hz) to fall to 2.27 Hz – on average, a 3.5% frequency shift for every metre of full-scale scour depth. Similar to Model 2, the natural frequency sensitivity of Model 3 showed reduced sensitivity with deeper local scour depths. For example, the frequency changed by 4.7% and 2.3%, respectively, for the first and second metre of local scour steps.

### 7.3.2 Global scour

The natural frequency of Model 3, measured with equivalent full-scale global scour depth of 2 m (2.21 Hz), was slightly lower than the natural frequency measured with 2 m of local scour (2.27 Hz). The global scour generated an overall frequency shift of 9.4%, which equates to 4.7% per metre of equivalent full-scale scour depth. Therefore, global scour shows 1.2% higher frequency sensitivity than local scour for every 1 m of full-scale scour. This higher frequency sensitivity is to be expected, as unscoured soil, slightly distant from the foundation, still provides a degree of increased pad foundation fixity, leading to higher stiffness and hence natural frequency for local scour relative to global scour.

## 7.4 Model 4: a monopile of Bridge 2

Figure 18 shows the full-scale natural frequency variation of the monopile. As expected, all of the natural frequency measurements at 60g fall between the fixed-base upper bound and 1g lower bound for all scour levels.

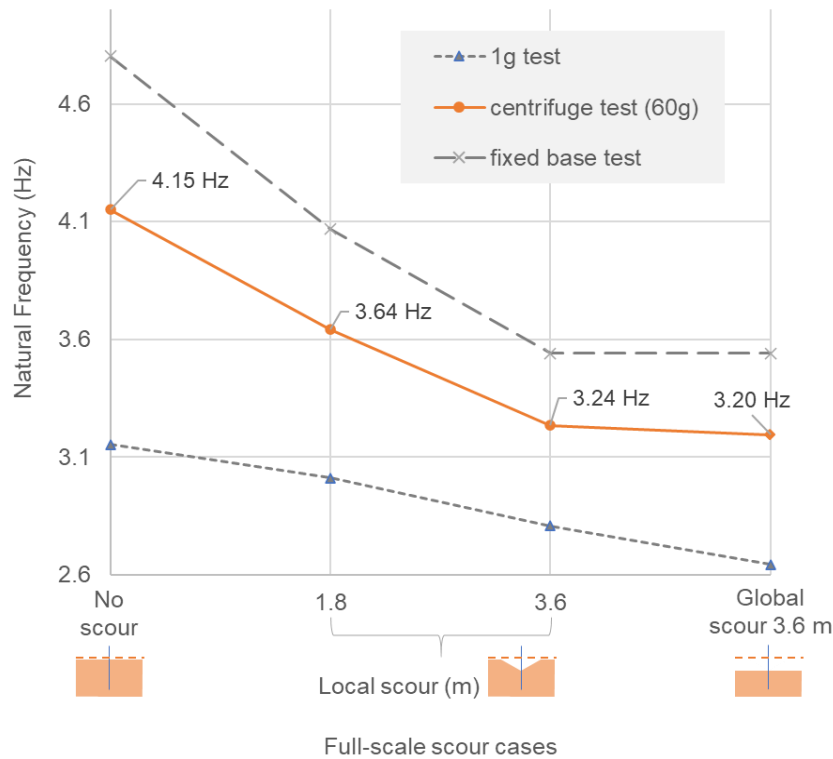

**Figure 18** *Full-scale natural frequency variation of monopile foundation*

### 7.4.1 Local scour

The full-scale local scour of 3.6 m around the monopile caused the fundamental natural frequency to fall from 4.15 Hz to 3.24 Hz. This represents a 22% change in natural frequency, equivalent to approximately 6.1% per metre of full-scale scour.

### 7.4.2 Global scour

Global scour of 3.6 m at full scale represents a natural frequency change of 22.9%, equivalent to approximately 6.4% per metre of scour. Global scour of 3.6 m caused the natural frequency to reduce slightly more (0.04 Hz) than the local scour of 3.6 m. However, the difference between the global- and local-scour-induced frequency shifts was only 0.3% per metre of scour.

## 8 Discussion

All foundation types exhibited a measurable change in natural frequency due to scour. A summary of the measured sensitivities is shown in Figure 19. The natural frequency reduction for global scour was greater than for local scour. This finding indicates that some additional confining pressure may be provided to the underlying soil by the overlying unscoured soil surrounding a local scour hole.

Current numerical models of soil-bridge systems that aim to simulate scour are limited by the representation of the soil-foundation interface. These use Winkler spring models to represent the soil small-strain stiffness, and scour is simulated by simply removing springs down to the level of scour, without making any distinction between local and global scour [15, 54]. The findings of this experiment suggest that improved models could be developed, to model the observed differences in the soil-structure interaction associated with these two scour types.

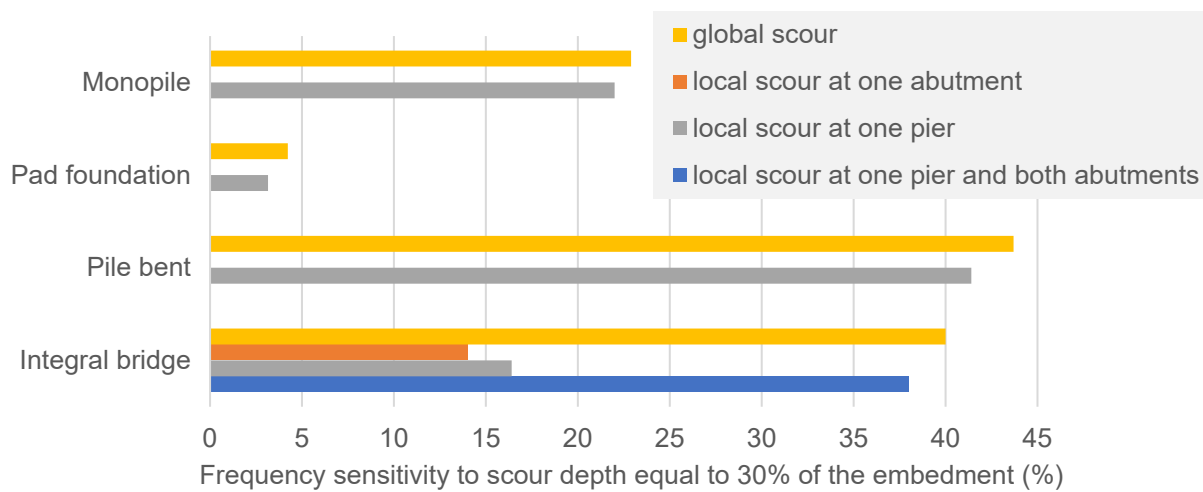

***Figure 19*** *Summary of frequency sensitivities at full scale*

Figure 19 shows the observed natural frequency sensitivities for a 30% loss of embedment (i.e. full-scale depths of 3.6 m for Models 1, 2 and 4 and 0.9 m for Model 3). It can be assumed that this is a significant loss of embedment of the bridge before any serious damage to the bridge. For this 30% loss of embedment, local scour at either the bridge pier or the abutment of the integral bridge model showed a 14–16% frequency reduction per metre of scour, while the pile bent model showed a 41% frequency reduction. The monopile showed a 22% frequency reduction, and the shallow pad foundation showed only a 3.2% reduction. Therefore, a natural frequency-based scour monitoring technique may have significant potential to indicate scour at bridges with deep piled foundations, but not for bridges with shallow pad foundations. Additionally, the results suggest that the frequency sensitivity to local scour reduces slightly with scour depth. This imply that there is a negative power relationship between natural frequency and exposed height (scour depth + initial height), similar to the behaviour expected in a simplified fixed cantilever representation of a bridge pier.

Some limitations of this research are worth noting. The fundamental vibration mode captured with Models 2, 3 and 4 (which have free ends with no restraints at the top) does not directly represent the fundamental mode of a simply supported bridge with the deck and bridge bearings of Bridges 2 and 3. With these additional stiffnesses and masses, the frequency sensitivity of a simply supported bridge may differ from the results reported here. However, the finding that deep foundations are more sensitive to scour than shallow foundations would still hold true in Bridges 2 and 3, since the introduction of the same deck arrangement can be expected to have a similar effect to the two foundation types. This aspect needs further investigation using numerical models of these foundation models, which can be calibrated using the experimental results observed here for local and global scour as explained in Kariyawasam (2020) [55].

Another limitation is that the soil model in the centrifuge experiment was of uniform density, whereas, in practice, density may vary with depth in different layers of soil. It is not clear to what extent this may be significant, especially in the case of piles where the pile-head stiffness is dominated by the soil restraint around the pile head (near the ground level), rather than the restraint offered at depth. A gradual reduction in frequency sensitivity with scour depth may not be noticed with a real bridge on layered soil. This aspect needs further investigation, with enhanced models to represent layered soils.

The centrifuge test was conducted in dry soil with no water present but the saturated or partially saturated soil present in a typical field-scale bridge may slightly change the frequency sensitivities to scour. However, such a change is unlikely to be significant, as previous numerical and experimental research suggests that the effect of water is negligible on the natural frequency sensitivity to scour of typical stiff bridge piers [24, 56, 57]. Further experimental research is required to study the natural frequency sensitivity of bridges in clay and saturated sand layers.

# 9   Conclusions

This paper has described the development of a centrifuge testing methodology to measure, for the first time, natural frequency sensitivity to local and global scour at the foundations of different bridge types. A scaled-down two-span integral bridge model and standalone foundations of two simply supported bridges were constructed in the sand with 66% relative density. These structural models were excited with a piezoelectric actuator and an automatic modal hammer developed for this test. The automatic modal hammer was found to have provided better impulsive excitations than the piezoelectric actuator.

The fundamental natural frequency of the integral bridge reduced by up to 40 % due to scour depths equal to approximately one third loss of pile embedment. These sensitivities are significantly higher than the reported natural frequency sensitivities to considerable structural damage and environmental sensitivities (0.4–7%) [9–14]. Thus, there is significant potential for natural frequency to be an indicator of integral bridges with piled foundations. The integral bridge considered here was a flexible support abutment, which has minimum interaction with the abutment backfill. In other types of integral bridge with abutment backfill abutment interaction, reliable natural frequencies may not be observed if the backfill interaction has significant changes with time.

The fundamental natural frequencies of standalone foundation models produced the highest sensitivity to a one-third loss of embedment for the monopile (44%), the second highest for the monopile (23%) and the lowest for the shallow foundation (4%). While these fundamental modes are not directly representative of the simply supported bridges with deck and bearings interactions, they do show that deep piled foundations have significantly greater natural frequency sensitivities than shallow foundations. Once the bridge deck is included to these standalone foundation models, the natural frequency sensitivities to scour could be lower than reported here since then the loss of soil-structure interaction because of scour becomes only a fraction of the overall stiffness provided by bridge deck, bearings and soil. Therefore, natural frequency is unlikely to be a reliable indicator for a simply supported bridge with shallow foundations. These experimental results for standalone foundations could be extrapolated to the simply supported bridges using numerical modelling in future studies.

The natural frequency reduction due to global scour was greater than for local scour; however, the difference was only 1–2% for one-third loss of embedment. Numerical modelling techniques could be developed in future studies to simulate these different effects of local and global scour. The frequency sensitivity to local scour tended to reduce as the scour hole deepened, so natural frequency as a measure of the progression of local scour may be more suitable for a bridge with no existing scour than for a bridge with severe local scour already present.

This study has demonstrated that natural frequency has significant potential as an indicator of scour for bridges with deep foundation/integral bridge decks. It is recommended that future studies consider further the influence of soil layering, as well as the likely frequency shifts due to other environmental factors such as temperature and water level, such that a detailed methodology for scour monitoring may be developed implementing this technique in practice.

## 10  Acknowledgements

We are grateful to the Cambridge Centre for Smart Infrastructure and Construction (Innovate UK grant reference number 920035 and EPSRC grant no EP/N021614/1), and the Laing O'Rourke Centre for Construction Engineering and Technology, for the funding provided for this experimental programme. The PhD funding provided by the Gates Cambridge Trust (Grant number OPP1144), founded by the Bill and Melinda Gates Foundation, is also gratefully appreciated. The valuable support provided by Tom Williams during the planning stage of this project is acknowledged. The research facilities provided by the Schofield Centre are gratefully acknowledged. We would like to acknowledge the initial development of the air hammer set-up by Peter Kirkwood and the piezoelectric actuator set-up by Jing Dong for testing monopiles. We thank Deryck Chan and Jad Boksmati for the valuable guidance provided to run the centrifuge test. We would also like to thank Geoff Eichhorn, Ahmed Alagha and Thejesh Garala for their guidance during the model preparation. Finally, thanks to Scofield Centre senor technicians Kristian Pether, Chris McGinnie, Mark Smith, John Chandler and David Layfield for their assistance at different stages of the experiment. Data supporting this paper is available from the University of Cambridge Open Data repository [58].

# 11 References

1. CIRIA (2015) Manual on scour at bridges and other hydraulic structures, 2nd edn. Department for Transport, London

2. Wardhana K, Hadipriono FC (2003) Analysis of recent bridge failures in the United States. J Perform Constr Facil 17:144–150. https://doi.org/Doi 10.1061/(Asce)0887-3828(2003)17:3(144)

3. Shirole AM, Holt RC (1991) Planning for a comprehensive bridge safety assurance program. Transp Res Rec No 1290 1:39–50

4. Fisher M, Atamturktur S, Khan AA (2013) A novel vibration-based monitoring technique for bridge pier and abutment scour. Struct Heal Monit 12:114–125. https://doi.org/10.1177/1475921713476332

5. Amirmojahedi M, Akib S, Basser H (2016) Methods for monitoring scour from large-diameter heat probe tests. Struct Heal Monit 15:. https://doi.org/10.1177/1475921715620004

6. Tang F, Chen Y, Li Z, et al (2019) Characterization and field validation of smart rocks for bridge scour monitoring. Struct Heal Monit 18:. https://doi.org/10.1177/1475921718824944

7. Zarafshan A, Iranmanesh A, Ansari F (2012) Vibration-Based Method and Sensor for Monitoring of Bridge Scour. J Bridg Eng 17:829–838. https://doi.org/10.1061/(ASCE)BE.1943-5592.0000362

8. Crotti G, Cigada A (2019) Scour at river bridge piers : real-time vulnerability assessment through the continuous monitoring of a bridge over the river Po, Italy. J Civ Struct Heal Monit 9:513–528. https://doi.org/10.1007/s13349-019-00348-5

9. Peeters B, Roeck G De (2001) One-year monitoring of the Z24-Bridge : environmental effects versus damage events. Earthq Engng Struct Dyn 30:149–171

10. Kim JT, Yun C-B, Yi J-H (2003) Temperature Effects on Modal Properties and Damage Detection in Plate-Girder Bridges. KSCE J Civ Eng 7:725–733. https://doi.org/10.1007/BF02829141

11. Koo KY, Brownjohn DI, Cole R (2010) Structural health monitoring of the Tamar Suspension. Struct Control Heal Monit 20:609–625. https://doi.org/10.1002/stc.1481

12. Caetano E, Magalha F (2009) Online automatic identification of the modal parameters of a long span arch bridge. Mech Syst Signal Process 23:316–329. https://doi.org/10.1016/j.ymssp.2008.05.003

13. Döhler M, Hille F, Mevel L, Rücker W (2014) Structural health monitoring with statistical methods during progressive damage test of S101 Bridge. Eng Struct 69:183–193. https://doi.org/10.1016/j.engstruct.2014.03.010

14. Farrar CR., Baker WE., Bell TM., et al (1994) Dynamic characterization and damage detection in the I-40 bridge over the Rio Grande

15. Prendergast, Hester D, Gavin K (2016) Determining the presence of scour around bridge foundations using vehicle-induced vibrations. J Bridg Eng 21:1–14. https://doi.org/10.1061/(ASCE)BE.1943-5592.0000931

16. Klinga J V., Alipour A (2015) Assessment of structural integrity of bridges under extreme scour conditions. Eng Struct 82:55–71. https://doi.org/10.1016/j.engstruct.2014.07.021

17. Prendergast LJ, Gavin K, Hester D (2017) Isolating the location of scour-induced stiffness loss in bridges using local modal behaviour. J Civ Struct Heal Monit 7:483–503. https://doi.org/10.1007/s13349-017-0238-3

18. Ko YY, Lee WF, Chang WK, et al (2010) Scour Evaluation of Bridge Foundations Using Vibration Measurement. In: 5th International Conference on Scour and Erosion (ICSE-5). ASCE, CA,

USA, pp 884–293

19. Shinoda M, Haya H, Murata S (2008) Nondestructive Evaluation of Railway Bridge Substructures By PERCUSSION TEST. Fourth Int Conf Scour Eros 285–290

20. Kariyawasam K, Fidler P, Talbot J, Middleton C (2019) Field deployment of an ambient vibration-based scour monitoring system at baildon bridge, UK. In: Proceedings of the International Conference on Smart Infrastructure and Construction. ICE UK, Cambridge, UK, pp 711–719. https://doi.org/10.1680/icsic.64669.711

21. Kariyawasam K, Fidler P, Talbot J, Middleton C (2019) Field assessment of ambient vibration-based bridge scour detection. In: Structural Health Monitoring. Stanford, CA, USA, pp 374–38. https://doi.org/10.12783/shm2019/32137

22. Kariyawasam K, Middleton C, Talbot J, et al (2021) On the potential for using bridge natural frequencies to detect scour: an experimental study with a field sensor deployment and geotechnical centrifuge modelling. In: IABSE Congress – Resilient technologies for sustainable infrastructure. IABSE, Christchurch, New Zealand

23. Boujia N, Schmidt F, Siegert D, et al (2017) Modelling of a bridge pier subjected to scour. Procedia Eng 199:2925–2930. https://doi.org/10.1016/j.proeng.2017.09.343

24. Prendergast LJ, Hester D, Gavin K, O'Sullivan JJ (2013) An investigation of the changes in the natural frequency of a pile affected by scour. J Sound Vib 332:6685–6702. https://doi.org/10.1016/j.jsv.2013.08.020

25. Madabhushi G (2014) Centrifuge Modelling for Civil Engineers. CRC Press, Boca Raton, FL

26. Garala TK, Madabhushi G (2020) Experimental investigation of kinematic pile bending in layered soils using Experimental investigation of kinematic pile bending in layered soils using dynamic centrifuge modelling. Geotechnique. https://doi.org/10.1680/jgeot.19.P.185

27. Futai MM, Dong J, Haigh SK, Madabhushi SPG (2018) Dynamic response of monopiles in sand using centrifuge modelling. Soil Dyn Earthq Eng 115:90–103. https://doi.org/10.1016/j.soildyn.2018.08.007

28. Sumer BM, Bundgaard K, Fredsøe J (2005) Global and Local Scour at Pile Groups. Int J Offshore Polar Eng 15:204–209

29. Mohamed A (2012) Chapter 1: Introduction to Offshore Structures. In: Offshore Structures : design. Elsevier, Waltham, USA, p 627

30. Seed HB, Idriss IM (1970) Soil Moduli and damping factors for dynamic response analyses Report EERC 70-10. CA, USA

31. Schofield AN (1980) Cambridge Geotechnical Centrifuge Operations. Geotechnique 30:227–268

32. Highways Agency (2001) The assessmenet of highway bridges and structures. In: Design Manual for Roads and Bridges, Volume 3, Section 4, Part 3 BD 21/01

33. Kaundinya I, Heimbecher F (2011) Identification and Classification of European Bridge and Tunnel Types. IABSE–IASS Symp

34. Steel Construction Institute (2018) Integral bridges. https://www.steelconstruction.info/Integral_bridges

35. Lee GC, Mohan SB, Huang C, Fard BN (2013) A Study of U.S. Bridge Failures (1980-2012)

36. O'Brien EJ, Keogh DL (1999) Bridge Deck Analysis, 1st ed. E & FN Spon, London

37. Lin C, Han J, Bennett C, Parsons RL (2013) Case History Analysis of Bridge Failures due to Scour. In: International Symposium of Climatic Effects on Pavement and Geotechnical Infrastructure. pp 204–216

38. State of Wisconsin Department of Transport (2018) WisDOT Bridge Manual Chapter 11 –

Foundation Support

39. Concast Precast Ltd (2009) Precast Concrete Beam Technical Manual

40. European Standard (2002) Eurocode 1: Actions on structures – Part 1-1: General actions – Densities, self-weight, imposed loads for buildings. Brussels

41. European Standard (2004) Eurocode 2: Design of concrete structures – Part 1-1: General rules and rules for buildings. Brussels

42. Teymur B, Madabhushi SPG (2003) Experimental study of boundary effects in dynamic centrifuge modelling. Géotechnique 53:655–663. https://doi.org/10.1680/geot.2003.53.7.655

43. Aalco (2019) Aluminium Alloy 6082 – T6~T651 Plate. http://www.aalco.co.uk/datasheets/Aluminium-Alloy_6082-T6~T651_148.ashx

44. Aalco (2019) Aluminium Alloy 6061 – T6 Extrusions. http://www.aalco.co.uk/datasheets/Aluminium-Alloy-6061-T6-Extrusions_145.ashx

45. DJB Instruments (2018) A / 23 / TS Piezoelectric Accelerometer

46. Analog Devices (2010) Single-Axis, High-g, iMEMS® Accelerometers ADXL78

47. Analog Devices (2017) Low Noise , High Frequency MEMS Accelerometers ADXL 1001/ADXL 1002

48. Madabhushi S, Houghton N, Haigh S (2006) A new automatic sand pourer for model preparation at University of Cambridge. Phys Model Geotech –- 6th ICPMG '06 217–222

49. Cedrat Technologies (2014) APA400MML datasheet

50. Hafez YI (2016) Mathematical Modeling of Local Scour at Slender and Wide Bridge Piers. J Fluids 2016:. https://doi.org/http://dx.doi.org/10.1155/2016/4835253

51. Brincker R, Anderson P, Jacobsen N-J (2007) Automated Frequency Domain Decomposition for Operational Modal Analysis. IMAC-XXV Conf Expo Struct Dyn

52. Brincker R, Zhang L, Andersen P (2001) Modal identification of output-only systems using frequency domain decomposition. Smart Mater Struct 10:441–445

53. Brandt A, Brincker R (2010) Impact Excitation Processing for Improved Frequency Response Quality. Proc IMAC-XXVIII

54. Bao T, Leo Z, Bird K (2019) Influence of soil characteristics on natural frequency-based bridge scour detection. J Sound Vib 446:195–210. https://doi.org/10.1016/j.jsv.2019.01.040

55. Kariyawasam KKGKD (2020) A vibration-based bridge scour monitoring technique (unpublished doctoral thesis). University of Cambridge, UK.

56. Boujia N, Schmidt F, Chevalier C, et al (2019) Effect of Scour on the Natural Frequency Responses of Bridge Piers : Development of a Scour Depth Sensor. infrastructures 4:. https://doi.org/10.3390/infrastructures4020021

57. Ju SH (2013) Determination of scoured bridge natural frequencies with soil-structure interaction. Soil Dyn Earthq Eng 55:247–254. https://doi.org/10.1016/j.soildyn.2013.09.015

58. Kariyawasam K, Middleton CR, Haigh S, et al (2020) Data supporting "Assessment of bridge natural frequency as an indicator of scour using centrifuge modelling." https://doi.org/10.17863/CAM.54904